\documentstyle[aps,preprint,tighten,epsf]{revtex}
\newcommand{\bfi}[1]{\mbox{\boldmath $#1$}}
\newcommand{\bfis}[1]{\mbox{\boldmath ${\scriptstyle #1}$}}
\begin{document}
\title{The optical potential of $^{6}$He in the eikonal approximation}
\author{B. Abu-Ibrahim$^{1,2}$ and Y. Suzuki$^1$}
\address{
$^1$ Department of Physics, Niigata University, Niigata 950-2181, Japan\\
$^2$ Department of Physics, Cairo University, Giza 12613, Egypt }
\date{\today}
\maketitle

\begin{abstract}
{The new data of the elastic scattering of 
$^{6}$He+$^{12}$C at about 40 MeV/nucleon
are analyzed in the eikonal approximation. 
The $^{6}$He+$^{12}$C phase-shift function is 
evaluated completely without any {\it ad hoc} 
assumption by a Monte Carlo integration, which makes it 
possible to use a realistic 6-nucleon wave function 
for a halo nucleus $^{6}$He. 
The effect of the breakup of $^6$He on the elastic differential cross 
sections as well as the optical potential 
is studied at different energies from 40 to 800 MeV/nucleon.\\
PACS number(s): 24.10.-i; 21.60.Ka; 25.60.Bx; 25.10.+s}

\end{abstract}

\section{Introduction}
A measurement of elastic scatterings of a projectile nucleus by 
a target nucleus provides unique information on the nuclear 
density distribution as well as the optical potential 
between them. The measurements of the angular distribution of 
purely elastic scattering for unstable nuclei are, however, 
very few so far because the separation of elastic and 
inelastic scatterings is in general difficult. Very recently 
the elastic scattering of $^6$He on $^{12}$C  
has been measured at 38.3 MeV/nucleon 
up to about 20$^{\circ}$ in the center-of-mass (c.m.) frame~\cite{lapoux}, 
which covers considerably wider angles than the previous measurement  
at 41.6 MeV/nucleon~\cite{alkhalili}. 
The new data will give us better opportunity to study the interaction 
potential between $^6$He and $^{12}$C. 

The nucleus $^{6}$He has attracted much experimental and theoretical 
interest as a two-neutron halo nucleus: 
It breaks to $^4$He+$n$+$n$ by the small energy input of 0.975 MeV. 
It is Borromean as the three-body system of $^4$He+$n$+$n$. At the present 
stage of experimental precision, the $\alpha$+$n$+$n$ model appears to be 
entirely satisfactory~\cite{tandf}. More sophisticated 
approach may be called for in some cases. 
For example, the breaking up of the 
$\alpha$-cluster is necessary to deeply understand 
the binding mechanism of $^6$He~\cite{arai}. 

The weak binding of a halo nucleus implies 
that it can easily decay to its constituents. 
Its ground state, lying close to the decay threshold, has 
a strong coupling to continuum states or resonances 
during the interaction 
with a target. This requires a special treatment of 
the coupling effect on the interaction potential for the halo 
projectile. 

A folding potential derived from an appropriate effective nuclear force, 
combined with a phenomenological imaginary term, has been successful 
for describing the elastic scattering of various systems~\cite{SL79}. 
A substantial 
reduction of the real strength of the folding potential has, however, been 
found necessary for weakly bound projectiles like $^{6,7}$Li and 
$^9$Be~\cite{SL79,SYK86}. This effect is explained by coupled-channel 
calculations in which projectile excitations are 
explicitly taken into account. The excitation 
of a weakly bound projectile into the continuum 
states has an important influence on the elastic scattering, and 
in the optical model this has to be mocked up by a substantial 
reduction of the depth of the real potential term. The difference
from the folding potential is called a dynamic polarization potential.

Since a halo nucleus is an extreme case of a weakly bound projectile, 
a similar effect is expected for halo-nucleus projectiles, thus 
the applicability of the folding model to halo nuclei
is questionable. In fact, the double-folding model using a 
realistic density-dependent nucleon-nucleon (NN) 
interaction, combined with the imaginary part  
taken in the conventional Woods-Saxon form, fails to 
reproduce the measured 
$^{6}$He+$^{12}$C elastic differential cross sections 
over the whole angular range~\cite{lapoux}. 

The inclusion of continuum states is performed by a 
continuum-discretized coupled-channel method, in which 
the continuum states of the projectile 
are approximated by a set of discrete states and 
each of them couples with the relative motion between the 
projectile and the target.  
For the two-neutron halo nucleus $^6$He the continuum states are 
three-body states, so in the continuum-discretized coupled-channel 
approach to the $^6$He+$^{12}$C scattering
one must solve a four-body Schr\"odinger equation, which, 
even if possible, will become very 
computer-time consuming. It is certainly impossible 
at present if one wants to 
consider the effect of, e.g., the distortion of the 
$\alpha$-core. 

A way out to overcome this difficulty is to use the eikonal 
approximation for composite-particle 
scattering~\cite{yabana92,alkhalili95,bertsch,bonaccorso}, 
which is known as the Glauber theory~\cite{Glauber} when all the 
constituents of the colliding nuclei are treated on an 
equal footing. We will show that 
a full Glauber calculation does not necessarily reproduce 
the new elastic scattering data but an effective phase calculated 
from a nucleon-target (NT) optical potential instead improves 
the description of the scattering. The utility of the approach has 
been tested even at low energies around 100 MeV/nucleon~\cite{badawy02}. 
With the use of the eikonal approximation 
it is straightforward to obtain the dynamic polarization potential. 
The purpose of this paper is to show that 
the new $^{6}$He+$^{12}$C data can substantially be understood 
in the eikonal approximation without introducing any {\it ad hoc} 
assumptions and in addition to study the energy dependence of the 
optical potential of the $^{6}$He+$^{12}$C system. We stress the 
characteristic points and the strength of the present approach: 
Firstly, a nucleon-$^{12}$C 
optical potential is used as a basic input of our calculation. 
Secondly, $^{6}$He is described with a realistic (6-nucleon) 
wave function which is obtained by variational Monte Carlo (VMC) 
calculation~\cite{pudliner}. By this the calculation of Ref.~\cite{badawy02} 
is further extended to directly use the wave function itself of $^6$He (
but not just the density). Thirdly, the scattering amplitude 
is calculated with the help of a Monte Carlo integration 
to all orders in the eikonal approximation. 

In sect.~\ref{eikonal} the scattering amplitude for composite-particle 
scatterings is briefly introduced 
in the eikonal approximation. In sect.~\ref{effective} an effective 
phase-shift function is defined by considering a target nucleus 
as a scatterer. 
The $^6$He+$^{12}$C elastic scattering at 40 MeV/nucleon is analyzed and 
the optical potential including the breakup of $^6$He is derived using 
the phase-shift function. 
In sect.~\ref{dpp} the energy dependence of the dynamic polarization 
potential is studied. A summary is given in sect.~\ref{summary}. 

\section{Eikonal approximation for composite-particle scattering}
\label{eikonal}
For a scattering problem of composite particles, the exact calculation 
of the scattering amplitude is extremely difficult. Appropriate 
approximations are indispensable. When the incident energy is high 
or the incident wave number $K$ is large enough, 
the relative motion between the projectile and the target 
oscillates rapidly and its deviation from a plane wave is expected to be 
small. This leads to the eikonal approximation. The applicability 
of the eikonal approximation may be tested by a 
comparison to a fully quantum-mechanical calculation for a simple 
potential scattering problem. For this aim we take a single-folding 
potential for elastic scatterings of $^6$He on $^{12}$C at 40 MeV/nucleon. 
(The detail of this potential will be explained later.)  It is found that 
the elastic differential cross section calculated in the 
eikonal approximation is by at most 15\% larger than the exact one  
up to 30$^{\circ}$ excepting around the cross section minimum. 
The discrepancy grows to about five times at 60$^{\circ}$. 
Thus the eikonal approximation works reasonably well for angles smaller 
than 30$^{\circ}$ at this energy. 
See also Ref.~\cite{atb} for the study of the accuracy 
of the eikonal calculations. 

A further approximation 
called an adiabatic approximation is needed to derive a simple, 
tractable expression for the scattering amplitude. In this 
approximation the excitation energies of the colliding nuclei 
are neglected. Under these approximations 
the scattering amplitude is given by

\begin{equation}
f_{\alpha \beta}(\theta, \phi)={i K \over 2\pi}\int d{\bfi b}\, 
{\rm e}^{-i{\bfis q}\cdot{\bfis b}}\langle \psi_{\alpha}^{(\rm P)}
\psi_{\beta}^{(\rm T)}| 1-{\rm e}^{i\sum_{i\in {\rm P}}
\sum_{j\in {\rm T}}
\chi_{\rm NN}({\bfis b}+{\bfis s}_i^{(\rm P)}-{\bfis s}_j^{(\rm T)})}
|\psi_{0}^{(\rm P)}\psi_{0}^{(\rm T)} \rangle,
\label{scatamp}
\end{equation}
where {\bfi q} is 
the momentum transferred from the target to the projectile, 
{\bfi b} a two-dimensional impact-parameter vector perpendicular to 
the $z$-direction, and e.g., ${\bfi s}_i^{(\rm P)}$ is the projection 
onto the $xy$-plane of the 
nucleon position vector relative to the projectile's c.m., 
${\bfi r}_i^{(\rm P)}\!-\!{\bfi R}_{\rm c.m.}^{(\rm P)}$. 
The wave function $\psi_{\alpha}^{(\rm P)}$ denotes the 
projectile's intrinsic state specified by a quantum number $\alpha$ with 
its c.m. part being dropped. ($\alpha=0$ 
stands for the ground state.) Similarly the target state is denoted 
$\psi_{\beta}^{(\rm T)}$. Thus $f_{00}$ stands for 
the elastic scattering amplitude.  See, for example, 
Refs.~\cite{Glauber,tandf} for more details. 

The phase-shift function $\chi_{\rm NN}$ in Eq.~(\ref{scatamp}) 
is a basic ingredient for the scattering amplitude. It describes 
the NN scattering and is related to the NN potential 
$V_{\rm NN}$ by 
\begin{equation}
\chi_{\rm NN}({\bfi b})=-{1\over \hbar v}
\int_{-\infty}^{+\infty}dz\, V_{\rm NN}({\bfi b}+z\hat{\bfi z}),
\end{equation}
where $v$ is the asymptotic velocity of the relative motion between 
the projectile and the target and $\hat{\bfi z}$ is a unit vector in 
the $z$-direction. The NN potential contains complicated 
spin-isospin dependence, so $\chi_{\rm NN}$ in general becomes an 
operator acting in that space. The use of such an operator in 
Eq.~(\ref{scatamp}) is extremely involved, and here it is treated as just 
a function by ignoring the spin-isospin dependence as is usually 
done. However, we distinguish protons and neutrons when necessary. 
The NN profile function $\Gamma_{\rm NN}({\bfi b})$ is introduced and 
often parametrized in the form
\begin{eqnarray}
\Gamma_{\rm NN}({\bfi b})&=&1-{\rm e}^{i\chi_{\rm NN}({\bfis b})}
\nonumber \\
&=&\frac{1-i\alpha}{2\pi}\,\omega\,\sigma_{\rm NN}\,{\rm e}^{
-\omega {\bfis b}^{2}}.
\label{profilefn}
\end{eqnarray}
Here $\sigma_{\rm NN}$ is the total NN cross section and the parameters 
$\alpha$ and $\omega$ are determined so as to fit the 
NN elastic differential cross section as well as the NN reaction cross section 
at relevant energy.  

To get the elastic scattering amplitude $f_{00}$ we 
need to calculate the phase-shift function $\chi({\bfi b})$:
\begin{equation}
{\rm e}^{i\chi({\bfis b})}=\langle\psi_{0}^{(\rm P)}\psi_{0}^{(\rm T)}
\vert\prod_{i\in {\rm P}}\prod_{j\in {\rm T}}
\left[1-\Gamma_{\rm NN}( {\bfi b}+{\bfi s}_{i}^{(\rm P)} - 
{\bfi s}_{j}^{(\rm T)} ) 
\right]\vert\psi_{0}^{(\rm P)}\psi_{0}^{(\rm T)}\rangle.
\label{psfunc}
\end{equation}
The above matrix element contains a multi-dimensional integration, 
which is obviously not easy to perform in general. Recently 
it has been demonstrated~\cite{varga02} that 
the phase-shift function can be evaluated by Monte Carlo method without 
approximation. The effectiveness of the method has been 
illustrated by 
several examples and will be used in this study as well. 

First we ask how well the Glauber model reproduces the 
elastic scattering data on $^{6}$He+$^{12}$C. 
The wave function used for $^{6}$He is the variational 
Monte Carlo (VMC) wave function~\cite{pudliner}. 
The VMC method starts with the construction
of a variational trial function of specified angular momentum,
parity and isospin, using products of two- and three-body correlation
operators acting on a fully antisymmetrized set of one-body basis states. 
This wave function is obtained by minimizing the energy of 
the nuclear Hamiltonian which includes 
realistic two- (Argonne $v_{18}$~\cite{wiringa95}) and three-body 
(Illinois IL2~\cite{pieper01}) interactions. The root mean square 
(r.m.s.) radii 
with the VMC wave function are 2.56, 1.96 and 2.81 fm for nucleon, proton and 
neutron, respectively. 
The $^{12}$C nucleus is not yet accessible in a realistic calculation 
and for the $^{12}$C
a three-$\alpha$ microscopic cluster-model
wave function is used: It gives the r.m.s. radius of 2.36 fm.
In this model the intrinsic wave function of
the $\alpha$-particles is a single shell-model Slater determinant and
the relative motion between the clusters is expressed in terms of linear
combinations of Gaussians. The combination coefficients
are determined variationally by solving the 12-nucleon
Schr\"odinger equation with an effective (Minnesota~\cite{minnesota}) 
two-nucleon interaction. 
Values of $\sigma_{\rm NN}$ and $\alpha$ defining $\Gamma_{\rm NN}$ 
are taken from literatures and the range parameter 
$\omega$ is determined by 
\begin{equation}
{1\over \omega}={1+\alpha^2 \over 8\pi}\sigma_{\rm NN},
\end{equation} 
which comes out from the condition that 
the NN total cross section be equal to 
the NN total elastic cross section. 
Results of calculation are shown in Fig.~\ref{6He12C.Glauber}: Dotted line is 
obtained by using $\sigma_{\rm NN}=13.5$ fm$^2$ and 
$\alpha=0.9$~\cite{lenzi}, while in dashed line 
different profile functions are 
used for np and pp (also nn) pairs, i.e., $\sigma_{\rm np}=21.8$ fm$^2$, 
$\alpha_{\rm np}=0.493$,  $\sigma_{\rm pp}=7$ fm$^2$, 
$\alpha_{\rm pp}=1.328$~\cite{hostachy}. 
Both calculations give similar cross sections up to 
14$^{\circ}$ and 
show some difference around the third and fourth minima. 
An interesting 
point is that the Glauber model reproduces the data only up to 
7$^{\circ}$ and beyond this angle gives  
much smaller cross sections than the measurement. Apparently this 
discrepancy indicates that at 40 MeV/nucleon multiple 
scatterings occurring in the composite-particle 
scattering are not quite well represented by the NN 
scattering determined in free space but receives some modifications by 
medium effects.

Though it does not reproduce the data well except for 
small angles, the Glauber model can include breakup effects of 
the projectile and the target. Before going to the next section, 
we attempt to assess the importance of the breakup effect of $^6$He in the  
elastic scattering. For this purpose we show in Fig.~\ref{6He12C.Glauber} the 
differential cross sections (solid line) 
calculated fully quantum-mechanically with the 
single-folding potential $U_{\rm f}(R)$
\begin{equation}
U_{\rm f}(R)=\int d{\bfi r}\, \rho^{(\rm P)}({\bfi r})
V_{\rm NT}({\bfi R}+{\bfi r})
\end{equation} 
with the projectile density
\begin{equation}
\rho^{(\rm P)}({\bfi r})
=\langle\psi_{0}^{(\rm P)}\vert \sum_{i\in {\rm P}}
\delta({\bfi r}_i^{(\rm P)}-{\bfi R}_{\rm c.m.}^{(\rm P)}-{\bfi r})
\vert \psi_{0}^{(\rm P)} \rangle ,
\end{equation}
where ${\bfi R}$ is the distance vector between the projectile and 
the target and $V_{\rm NT}$ is an NT optical potential. 
This single-folding model gives both real and 
imaginary potentials automatically. The folding model is found to 
overestimate the measured cross sections beyond 6$^{\circ}$, which 
is in contrast 
to the full Glauber-model calculation. It also cannot 
reproduce simultaneously the first deep 
minimum and the third maximum as noted in Ref.~\cite{lapoux}, 
where the double-folding model is used to generate the 
$^6$He-$^{12}$C potential and the imaginary part was adjusted 
to fit the data. 
The overestimation of the single-folding model is due to 
neglecting the breakup effect of $^6$He. This will be discussed 
later in more detail. 

\section{Effective eikonal phase}
\label{effective}
In the previous section we have observed that the Glauber model 
with the NN profile function does not necessarily reproduce the 
data. Since the NN profile function is chosen to be consistent 
with the data of the elementary NN scattering in free space, 
this suggests that an appropriate effective 
interaction has to be employed in the calculation. Instead of 
using such an effective interaction like $G$-matrix, we 
proposed a simple, 
practical approach~\cite{badawy00,badawy02} to 
composite-particle scatterings 
by considering the target just a scatterer and taking 
an NT scattering as an elementary vehicle. 

In this formalism various effects
such as the Fermi motion and the Pauli-blocking, etc., 
would be included to some extent through the NT amplitude 
determined from an NT optical potential. 
The effective optical phase-shift function is thus given by
\begin{eqnarray}
{\rm e}^{i\widetilde{\chi}({\bfis b})}
&=&\langle\psi_{0}^{(\rm P)}\vert{\rm e}^{i\sum_{i\in {\rm P}}
\chi_{\rm NT}({\bfis b}+{\bfis s}_i^{(\rm P)})}|\psi_{0}^{(\rm P)} \rangle 
\nonumber \\
&=&\langle\psi_{0}^{(\rm P)}\vert\prod_{i\in {\rm P}}
\left[1-\Gamma_{\rm NT}({\bfi b}+ {\bfi s}_{i}^{(\rm P)} ) \right]\vert
\psi_{0}^{(\rm P)}\rangle,
\label{gnt}
\end{eqnarray}
where $\Gamma_{\rm NT}=1-{\rm e}^{i\chi_{\rm NT}}$ is 
related to $V_{\rm NT}$ through  
\begin{equation}
\chi_{\rm NT}({\bfi b})=-\frac{1}{\hbar v}
\int_{-\infty}^{\infty} dz\,V_{\rm NT}({\bfi b}+ z\hat{\bfi z} ).
\label{eik}
\end{equation}
The examples of calculation shown in Ref.~\cite{badawy00} 
were obtained by approximating Eq.~(\ref{gnt}) by
\begin{equation}
{\rm e}^{i\widetilde{\chi}_{\rm OLA}({\bfis b})}
={\rm exp}\left\{-\int d{\bfi r}\,\rho^{(\rm P)}({\bfi r})
\Gamma_{\rm NT}({\bfi b}+{\bfi s})\right\},
\end{equation}
where ${\bfi s}$ is the projection of ${\bfi r}$ onto 
the $xy$-plane. 
In the present study we will not use this approximation but perform a 
virtually exact phase-shift calculation 
by a Monte Carlo method in which $\vert \psi_0^{(\rm P)}\vert^2$ serves 
as a weight function for sampling configuration points.

It should be noted that the phase-shift function corresponding to the 
single-folding potential appears as the first term of a cumulant expansion 
of Eq.~(\ref{gnt}). To see this we note that the 
phase-shift function $\chi_{\rm f}({\bfi b})$ 
corresponding to $U_{\rm f}(R)$ is given by 
\begin{eqnarray}
\chi_{\rm f}({\bfi b})&=&-\frac{1}{\hbar v}
\int_{-\infty}^{\infty} dz\,U_{\rm f}({\bfi b}+ z\hat{\bfi z} ) 
\nonumber \\
&=&\int d{\bfi r}\, \rho^{(\rm P)}({\bfi r})
\chi_{\rm NT}({\bfi b}+{\bfi s}) 
\nonumber \\
&=&\langle\psi_{0}^{(\rm P)}\vert \sum_{i\in {\rm P}}
\chi_{\rm NT}({\bfi b}+{\bfi s}_i^{(\rm P)})\vert \psi_{0}^{(\rm P)} \rangle.
\label{chifold}
\end{eqnarray}
By substituting Eq.~(\ref{chifold}) into Eq.~(\ref{gnt}), 
we obtain the following expression 
\begin{equation}
{\rm e}^{i\widetilde{\chi}({\bfis b})}=
{\rm e}^{i{\chi}_{\rm f}({\bfis b})}
\langle\psi_{0}^{(\rm P)}\vert{\rm e}^{i\delta \chi}
\vert \psi_{0}^{(\rm P)} \rangle 
\label{psfvsfold}
\end{equation}
with
\begin{equation}
\delta \chi=\sum_{i\in {\rm P}}
\chi_{\rm NT}({\bfi b}+{\bfi s}_i^{(\rm P)})-{\chi}_{\rm f}({\bfi b}),
\end{equation}
which will be exploited when we discuss 
the dynamic polarization potential in sect.~\ref{dpp}.

We first consider $^{4}$He+$^{12}$C elastic scattering at 
40 MeV/nucleon. The study of this system will give us 
information on the $^{6}$He+$^{12}$C elastic 
scattering. The $^{4}$He+$^{12}$C phase-shift 
function $\widetilde{\chi}$ is calculated by using the 
VMC wave function for $^{4}$He and three different 
p-$^{12}$C optical potentials taken from 
Refs.~\cite{fannon,becchetti,rapaport}. The 
spin-orbit component of the optical potential is 
ignored in the calculation. All of the optical 
potentials reproduce reasonably well the p+$^{12}$C elastic 
scattering data~\cite{blumberg} up to about 50$^{\circ}$, 
but show some deviation from the experimental data beyond that angle. 
The p+$^{12}$C and $^6$He+$^{12}$C reaction cross sections calculated by 
these potentials are listed in Table~\ref{table1}. 
All the potentials have a similar real part except that 
Fannon {\it et al.} potential (the set of $V_0=-$47.2 MeV) 
is shorter-ranged and 
has deeper central strength than the others. For the imaginary part 
Fannon {\it et al.} potential is deeper than the others 
near the surface. 
A comparison with experiment~\cite{tb70} for $^{4}$He+$^{12}$C scattering is 
shown in Fig.~\ref{4He12C.eikonal}. The potential of Ref.~\cite{fannon} 
(solid line) reproduces the data quite well up to 15$^{\circ}$ and  
slightly underestimates the cross sections at larger angles, while 
the other potentials (dashed and dash-dotted lines) 
overestimate the cross section at 
the minimum around 10$^{\circ}$. To see the sensitivity 
of the cross section to wave functions, we repeated the calculation 
by using the single harmonic-oscillator shell-model 
wave function for $^4$He with its c.m. part being dropped. 
The oscillator parameter was set to 
reproduce the same r.m.s. radius as that of the VMC wave function (1.46 fm).
The deviation from the VMC case was very small at this energy. 
As shown in Ref.~\cite{varga02}, different predictions of simple 
model and realistic wave functions can be seen in cross sections 
corresponding to much higher momentum transfer.  

One of the advantages 
of the present approach is that we rigorously took account of the 
c.m. problem, thus using the intrinsic coordinate 
${\bfi s}_i^{(\rm P)}$ measured from the c.m. of the projectile.
Several authors discussed the effect of the c.m. correlation 
when one calculates the Glauber amplitude~\cite{czyz,c.m.correl}. 
Dotted line in Fig.~\ref{4He12C.eikonal} indicates 
the cross section calculated with the same shell-model 
wave function but replacing ${\bfi s}_i^{(\rm P)}$ with the projection 
of ${\bfi r}_i^{(\rm P)}$ itself onto the $xy$-plane. Ignoring the 
c.m. correlation is found to cause a significant error 
for such a light nucleus as $^4$He.  

The $^{6}$He+$^{12}$C phase-shift function $\widetilde{\chi}$ 
is calculated by using the VMC wave function for $^{6}$He. 
The cross sections calculated with the three p-$^{12}$C optical 
potentials are displayed in Fig.~\ref{6He12C.eikonal}. Rapaport 
potential (solid line) and Becchetti and Greenlees potential 
(dashed line) appear to give a fair agreement with the 
experimental data though at angles less than 10$^{\circ}$ 
Fannon {\it et al.} potential (dotted line) better fits the data.  
The difference in the cross sections is 
mainly due to that of the imaginary part of the p-$^{12}$C 
potential. Though all the potentials give similar results of 
both the elastic differential cross section and 
the reaction cross section for p+$^{12}$C, the $^{6}$He+$^{12}$C 
cross sections are different beyond the second minimum. 
Solid lines in Figs.~\ref{6He12C.Glauber} and \ref{6He12C.eikonal} 
employ the same p-$^{12}$C 
optical potential~\cite{rapaport}. The difference is that the former 
is the folding-model calculation, while the latter 
is the eikonal calculation. The overestimation of the cross 
sections observed in the folding-model calculation is 
certainly improved by the eikonal calculation. 

The local, energy-dependent, optical potential for 
the projectile-target scattering can be 
constructed from the phase-shift function by~\cite{Glauber}
\begin{equation}
U(R)=\frac{\hbar v}{\pi}\frac{1}{R}\frac{d}{dR}
\int_{R}^{\infty}db\,{ b\frac{\widetilde{\chi}(b)}
{\sqrt{b^{2}- R^{2}}}}.
\label{pot}
\end{equation}
Once the potential $U(R)$ is constructed, we can solve a standard radial 
Schr\"odinger equation with $U(R)$ 
to obtain the exact partial-wave phase shift $\delta_l$. 
Provided the eikonal approximation is valid, we have 
\begin{equation}
\widetilde{\chi}(b) \approx 2\delta_l,\ \ \ \ \ 
bK=l+{1\over 2}.
\end{equation} 
Figure~\ref{6He12C.QM} shows the elastic differential 
cross sections calculated with $\delta_l$. The differential cross 
sections in Figs.~\ref{6He12C.eikonal} and \ref{6He12C.QM} are thus calculated 
with the same optical potential $U(R)$. The difference is that 
the eikonal approximation is used in Fig.~\ref{6He12C.eikonal}, whereas 
a partial-wave expansion for the scattering amplitude is used in 
Fig.~\ref{6He12C.QM}. As was mentioned in the beginning of sect.~\ref{eikonal},
the eikonal approximation is found to slightly overestimate 
the differential cross section compared to the exact value.

The difference between $U(R)$ and the single-folding 
potential $U_{\rm f}(R)$ is nothing but the dynamic polarization 
potential in the eikonal approximation: 

\begin{equation}
U_{\rm DPP}(R) = U(R)-U_{\rm f}(R).
\label{potser}
\end{equation} 

If the $^6$He wave function is factorized 
as a three-body wave function $\psi_{\rm FB}$ consisting of the $\alpha$-core 
and two-neutron parts, the effective phase-shift function of Eq.~(\ref{gnt}) 
may be approximated as follows:
\begin{equation}
{\rm e}^{i{\chi}_{\rm FB}({\bfis b})}=\langle\psi_{\rm FB}\vert
\left[1-\Gamma_{\rm CT}({\bfi b}+ {\bfi s}_{\rm C}) \right]\prod_{i=1,2}
\left[1-\Gamma_{\rm NT}({\bfi b}+ {\bfi s}_{i}) \right]\vert
\psi_{\rm FB}\rangle,
\end{equation}
where ${\bfi s}_{\rm C}$ is the projection onto the $xy$-plane 
of the core-c.m. position vector measured 
from the c.m. and the core-target profile function 
$\Gamma_{\rm CT}({\bfi b})$ is calculated from an $\alpha$-$^{12}$C 
optical potential. 
This approximation is sometimes called a few-body Glauber model. It is to be 
noted that even though the $\alpha$+$n$+$n$ model for $^6$He is 
satisfactory the factorization of the 6-nucleon wave function of $^6$He 
cannot in general be performed rigorously. Figure~\ref{6He12C.FBGlau} 
displays the differential cross section calculated with a three-body wave 
function where two-neutron halo wave function is generated from 
$0p_{1/2}$ orbit~\cite{skoyb}. The $\alpha$-$^{12}$C 
optical potential is taken from Ref.~\cite{tb70}. 
The cross section calculated by the few-body Glauber model 
is as good as that of the eikonal calculation using 
the 6-nucleon wave function (compare to Fig.~\ref{6He12C.eikonal}). However, 
it is found that the different 
p-$^{12}$C optical potentials lead to only a small difference in the cross 
section in the case of the few-body Glauber model, which is in contrast to the 
6-nucleon calculation. This is probably because 
the $\alpha$-$^{12}$C interaction dominates the whole 
phase-shift responsible for the scattering process. It seems that 
the approach based on the 6-nucleon 
eikonal calculation is more natural than the few-body approach 
particularly when we discuss the energy dependence of the 
optical potential for $^{6}$He.

\section{Dynamic polarization potential of $^{6}$He}
\label{dpp}
From Eq.~(\ref{psfvsfold}) we have the following relation 
\begin{eqnarray}
\widetilde{\chi}({\bfi b})&=&\chi_{\rm f}({\bfi b})-i{\rm ln}
\langle\psi_{0}^{(\rm P)}\vert{\rm e}^{i\delta \chi}
\vert \psi_{0}^{(\rm P)} \rangle \nonumber \\
&=& \chi_{\rm f}({\bfi b}) 
+\frac{i}{2}\langle (\delta \chi)^{2} \rangle
-\frac{1}{6}\langle (\delta \chi)^{3} \rangle
-\frac{i}{24}\langle (\delta \chi)^{4} \rangle
+\frac{i}{8}\langle (\delta \chi)^{2} \rangle^2
+\cdots
\label{psser}
\end{eqnarray}
Here e.g., $\langle (\delta \chi)^{2} \rangle =
\langle \psi_{0}^{(\rm P)}\vert (\delta \chi)^{2}
\vert \psi_{0}^{(\rm P)} \rangle $ and use is made of the property 
$\langle \delta \chi \rangle =0$. Substituting this 
cumulant expansion to Eq.~(\ref{pot}) 
we have the corresponding decomposition of the optical potential, and  
the first term in the expansion is nothing but 
the folding potential $U_{\rm f}$. Thus the dynamic polarization 
potential is contributed by those potentials that correspond to the 
terms $\langle (\delta \chi)^{n} \rangle \ (n=2,3,\ldots)$ 
arising from the breakup of the projectile. 

One can predict the sign of the dynamic polarization 
potential under certain conditions~\cite{yabana92,tandf}. 
Assuming that the $V_{\rm NT}$ 
has the form $V_{\rm NT}(R)\sim (V_0+iW_0)f(R)$, the sign 
of the term $\langle (\delta \chi)^{2} \rangle$ is 
determined by $(V_0+iW_0)^2$. 
If the dynamic polarization potential is contributed dominantly by the term 
${i \over 2}\langle (\delta \chi)^{2} \rangle$, its sign follows 
$-i(V_0+iW_0)^2$, so the real and imaginary parts of the dynamic 
polarization potential have the same sign as $V_0W_0$ and $W_0^2-V_0^2$, 
respectively. 

We compare in Fig.~\ref{6He12C.OM} the $^{6}$He+$^{12}$C 
single-folding potential (dashed line) with the optical 
potential calculated from the eikonal approximation (solid line). 
The breakup process produces a repulsive surface effect 
on the real part of the optical potential and increases 
a strength of the imaginary part reflecting the 
loss of the flux due to the elastic breakup process. 
Dash-dotted and dotted lines in the figure denote the optical potential 
and the single-folding potential for 
$^{4}$He+$^{12}$C scattering calculated at the same energy of 
40 MeV/nucleon. The breakup effect does not practically change the real part 
of the potential but moderately increases the strength of the 
imaginary part for the $^{4}$He+$^{12}$C case. The effect becomes 
very significant for the $^{6}$He+$^{12}$C case. In particular 
the imaginary part of the $^{6}$He-$^{12}$C optical potential becomes much 
longer-ranged and almost twice stronger than that of the 
$^{4}$He-$^{12}$C optical potential. 
   
Next we study the energy dependence of the breakup effect. 
Figure~\ref{6He12C.E-dep} shows calculated $^{6}$He+$^{12}$C 
elastic differential cross sections at different energies from 70 
MeV/nucleon to 800 MeV/nucleon. We used 
the p-$^{12}$C global optical potential~\cite{cooper}. 
The energy dependence of this potential is as follows: 
The real part at the center alters as $-32.3, \, -11.7,\, 16.2$ MeV, 
for $70,\, 200,\, 800$ MeV/nucleon, respectively, while the imaginary 
part as $-0.9,\, -12.0,\,-87.2$ MeV. 
Solid line in Fig.~\ref{6He12C.E-dep} denotes the eikonal calculation, 
while dashed line the folding-model calculation. 
Figure~\ref{6He12C.OM.E-dep} compares the real and 
imaginary parts of the $^{6}$He-$^{12}$C 
optical potential $U(R)$ determined from the phase-shift function 
with those of the single-folding potential $U_{\rm f}(R)$. 
The real part of the folding potential becomes shallower as the energy 
increases and turns out to be repulsive at 800 MeV/nucleon, 
whereas the imaginary part of the folding potential becomes 
deeper with the energy increasing. 
At 70 MeV/nucleon both of the real and imaginary parts of 
the p-$^{12}$C global optical potential is negative and $V_0^2 > W_0^2$, so 
the first term of the dynamic polarization potential 
has positive real and negative imaginary parts, respectively. 
This explains why the optical potential is less attractive 
and more absorptive than the single-folding potential at this energy. 
At 800 MeV/nucleon the global potential has 
positive real and negative imaginary parts with $W_0^2 > V_0^2$, so 
the first term of the dynamic polarization potential has 
a negative sign for the real part and a positive sign for the 
imaginary part, which also explains the change of the optical 
potential from the single-folding potential at this energy. 
The magnitude of the dynamic polarization potential at  800 MeV/nucleon 
is quite big: Quite large, negative imaginary potential of the 
folding potential is dramatically reduced to $-49.3$ MeV at the center.

\section{Summary}
\label{summary}
With the Monte Carlo integration we evaluated the scattering 
amplitude of the Glauber model without any {\it ad hoc} approximation. 
The advantage of this approach is that one can 
use accurate, sophisticated wave functions of colliding nuclei. 

We studied the elastic scattering of $^{6}$He+$^{12}$C at 40 MeV/nucleon 
in order to see the breakup effect of a weakly bound halo nucleus $^6$He. 
Comparing to the new $^{6}$He+$^{12}$C data which cover larger angles 
than the previous data, we found that the complete Galuber amplitude with 
use of a realistic VMC wave function for $^{6}$He reproduced experiment 
only at small angles but beyond the first minimum 
gave too small differential cross sections. Calculating the phase-shift 
function with a nucleon-$^{12}$C optical potential, we 
were able to reproduce the experimental data reasonably well. In this 
calculation the breakup effect of $^{6}$He was taken into account 
in the eikonal approximation.  

Through the calculated phase-shift function we constructed the 
$^{6}$He-$^{12}$C optical potential which contains the dynamic 
polarization potential due to the breakup effect of $^{6}$He. 
This optical potential has a longer range and a deeper depth than 
that of the $^{4}$He-$^{12}$C optical potential. The 
difference between the two potentials appears more strongly in the imaginary 
part. We studied the elastic scattering of $^{6}$He+$^{12}$C at 
different energies and discussed the energy dependence of the 
dynamic polarization potential from 40 to 800 MeV/nucleon. 
The effect of the breakup is so strong that e.g., the imaginary part 
of the dynamic polarization potential renders the central depth of the 
folding potential ($-314.1$ MeV) considerably shallow ($-49.3$ MeV). 

The approach presented here has the strength that it 
makes it possible to directly relate cross sections to relevant 
wave functions. We thus hope that the new radioactive beam facilities 
will produce experimental data covering larger angles. 

\section{Acknowledgements}
The authors thank K. Varga, V. Lapoux, B. C. Clark, N. Orr and 
J. Y. Hostachy for useful communications. One of the authors (B. A-I.) 
is supported by a JSPS Postdoctoral Fellowship for 
Foreign Researchers (No. P03023). This work was in part 
supported by a Grant-in-Aid for Scientific Research (No. 14540249) of the 
Ministry of Education, Science, Sports and Culture (Japan).

\begin{table}
\caption{Reaction cross sections in mb for p+$^{12}$C and $^{6}$He+$^{12}$C 
reactions predicted by different p-$^{12}$C optical potentials.}
\begin{center}
\begin{tabular}{lcc}
Optical potential        & p+$^{12}$C  &  $^{6}$He+$^{12}$C  \\
\hline
Fannon {\it et al.}~\protect\cite{fannon}   &  380    &   1193     \\
Becchetti and Greenlees~\protect\cite{becchetti} &  377   &   1147    \\
Rapaport~\protect\cite{rapaport}  &  396    &   1139    \\
\end{tabular}
\end{center}
\vspace*{-0.5cm}
The experimental p+$^{12}$C reaction cross section is 
371$\pm$9 mb~\protect\cite{carlson}.
\label{table1}
\end{table}

\begin{figure}
\caption{Elastic differential cross sections in Rutherford ratio
for $^{6}$He+$^{12}$C scattering at 40 MeV/nucleon. 
Solid line is the cross section calculated with a 
single-folding model in which the p-$^{12}$C optical 
potential~\protect\cite{rapaport} is folded with the 
VMC density~\protect\cite{pudliner}. Dotted and dashed lines are 
full Glauber-model calculations which employ the 
sets of $\Gamma_{NN}$ parameters as explained in text. 
The data are taken from Refs.~\protect\cite{lapoux,alkhalili}.}
\label{6He12C.Glauber}
\end{figure}

\begin{figure}
\caption{Elastic differential cross sections in Rutherford ratio
for $^{4}$He+$^{12}$C scattering at 40 MeV/nucleon. The VMC wave function for 
$^4$He~\protect\cite{pudliner} is used. Solid, dash-dotted and dashed 
lines denote the results with three different p-$^{12}$C optical 
potentials~\protect\cite{fannon,becchetti,rapaport}, respectively. 
Dotted line is the cross section calculated by using the single 
harmonic-oscillator shell-model wave function for $^4$He and the 
potential of Ref.~\protect\cite{fannon} but by ignoring the c.m. correlation 
as explained in text. The data are taken from Refs.~\protect\cite{tb70}. }
\label{4He12C.eikonal}
\end{figure}

\begin{figure}
\caption{Elastic differential cross sections in Rutherford ratio  
calculated in the eikonal approximation 
for $^{6}$He+$^{12}$C scattering at 40 MeV/nucleon. The VMC wave function for 
$^6$He~\protect\cite{pudliner} is used.  Dotted, dashed 
and solid lines denote the results with three different p-$^{12}$C optical 
potentials~\protect\cite{fannon,becchetti,rapaport}, respectively. }
\label{6He12C.eikonal}
\end{figure}

\begin{figure}
\caption{The same as in Fig.~\protect\ref{6He12C.eikonal} but by 
a partial-wave expansion for the scattering amplitude.}
\label{6He12C.QM}
\end{figure}

\begin{figure}
\caption{The same as in Fig.~\protect\ref{6He12C.eikonal} but by the 
few-body Glauber model. See text for the three-body wave function and 
the $\alpha$-$^{12}$C potential employed in the calculation.}
\label{6He12C.FBGlau}
\end{figure}

\begin{figure}
\caption{Real and imaginary parts of the $^{6}$He-$^{12}$C 
optical potential at 40 MeV/nucleon. 
Solid line includes the breakup effect of $^6$He, while dashed line 
is the single-folding potential. The difference of the two potentials 
is the dynamic polarization potential. The p-$^{12}$C optical 
potential~\protect\cite{rapaport} is used. Dash-dotted and dotted lines 
denote the optical and single-folding potentials for $^{4}$He-$^{12}$C 
at 40 MeV/nucleon, respectively. The VMC wave 
functions~\protect\cite{pudliner} are used for both $^6$He and $^4$He. } 
\label{6He12C.OM}
\end{figure}

\begin{figure}
\caption{Elastic differential cross sections in Rutherford ratio
for $^{6}$He+$^{12}$C scattering calculated at different energies. 
The p-$^{12}$C optical potential is taken from Ref.~\protect\cite{cooper}. 
The VMC wave function~\protect\cite{pudliner} is used for $^6$He. 
Solid line is the eikonal calculation, 
while dashed line is the single-folding model calculation.} 
\label{6He12C.E-dep}
\end{figure}

\begin{figure}
\caption{Real (a) and imaginary (b) parts of the $^{6}$He-$^{12}$C 
optical potential at different energies. 
Solid line includes the breakup effect of $^6$He, while dashed line 
is the single-folding potential. The p-$^{12}$C optical potential 
is taken from Ref.~\protect\cite{cooper}. The VMC wave 
function~\protect\cite{pudliner} is used for $^6$He. }
\label{6He12C.OM.E-dep}
\end{figure}


\begin{thebibliography}{99}
\bibitem{lapoux} V. Lapoux {\it et al.}, Phys. Rev.
C {\bf 66} (2002) 034608.
\bibitem{alkhalili} J. S. Al-Khalili {\it et al.}, Phys. Lett.
B {\bf 378} (1996) 45.
\bibitem{tandf} see, for example, 
Y. Suzuki, R. G. Lovas, K. Yabana and K. Varga, 
{\it Structure and Reactions of Light Exotic Nuclei} (Taylor \& Francis, 
London, 2003).
\bibitem{arai} K. Arai, Y. Suzuki and R. G. Lovas, Phys. Rev. C {\bf 59} 
(1999) 1432.
\bibitem{SL79} G. R. Satchler and W. G. Love, Phys. Rep. 55 (1979) 184.
\bibitem{SYK86} Y. Sakuragi, M. Yahiro and M. Kamimura, 
Prog. Theor. Phys. Suppl. No. 89 (1986) 136.
\bibitem{yabana92} K. Yabana, Y. Ogawa and Y. Suzuki,
Phys. Rev. C {\bf 45} (1992) 2909.
\bibitem{alkhalili95} J. S. Al-Khalili, I. J. Thompson and J. A. Tostevin.
Nucl. Phys. A {\bf 581} (1995) 331.
\bibitem{bertsch} G. F. Bertsch, K. Hencken and H. Esbensen,
Phys. Rev. C {\bf 57} (1998) 1366.
\bibitem{bonaccorso} A. Bonaccorso and D. M. Brink, Phys. Rev. C {\bf 58} 
(1998) 2864.
\bibitem{Glauber} R. J. Glauber, in {\it Lectures on Theoretical Physics}, 
edited by W. E. Brittin and L. C. Dunham (Interscience, New York, 1959), Vol.1, p.315.
\bibitem{badawy02} B. Abu-Ibrahim and Y. Suzuki, 
Nucl. Phys. A {\bf 706} (2002) 111.
\bibitem{pudliner} B. S. Pudliner, V. R. Pandharipande, J. Carlson,
S. C. Pieper and R. B. Wiringa, Phys. Rev. C {\bf 56} (1997) 1720.
\bibitem{atb} J. S. Al-Khalili, J. A. Tostevin and J. M. Brooke, Phys. Rev. 
C {\bf 55} (1997) R1018; J. M. Brooke, J. S. Al-Khalili and J. A. Tostevin, 
Phys. Rev. C {\bf 59} (1999) 1560.
\bibitem{varga02} K. Varga, S. C. Pieper, Y. Suzuki 
and R. B. Wiringa, Phys. Rev. C {\bf 66} (2002) 034611.
\bibitem{wiringa95} R. B. Wiringa, V. G. J. Stocks and R. Schiavilla, 
Phys. Rev. C {\bf 51} (1995) 38.
\bibitem{pieper01} S. C. Pieper, V. R. Pandharipande, R. B. Wiringa and 
J. Carlson, Phys. Rev. C {\bf 64} (2001) 014001.
\bibitem{minnesota} D. R. Thompson, M. LeMere and Y. C. Tang, 
Nucl. Phys. A {\bf 286} (1977) 53.
\bibitem{lenzi} S. M. Lenzi, A. Vitturi and F. Zurdi, 
Phys. Rev. C {\bf 40} (1989) 2114.
\bibitem{hostachy} J. Y. Hostachy, Th$\grave{\rm e}$se 
d'Etat, I.S.N. 87-65, Universit$\acute{\rm e}$ de Grenoble, unpublished. 
\bibitem{badawy00} B. Abu-Ibrahim and Y. Suzuki, Phys. Rev. C {\bf 61} 
(2000) 051601(R); {\it ibid.} {\bf 62} (2000) 034608.
\bibitem{fannon} J. A. Fannon, E. J. Burge, D. A. Smith and N. K. Ganguly, 
Nucl. Phys. A {\bf 97} (1967) 263.
\bibitem{becchetti} F. D. Becchetti and G. W. Greenlees,
Phys. Rev. {\bf 182} (1969) 1190.
\bibitem{rapaport} J. Rapaport, Phys. Rep. {\bf 87} (1982) 25.
\bibitem{blumberg} L. N. Blumberg, E. E. Gross, A. van der Woude, 
A. Zucker and R. H. Bassel, Phys. Rev. {\bf 147} (1966) 812.
\bibitem{tb70} B. Tatischeff and I. Brissaud, Nucl. Phys. A {\bf 155} 
(1970) 89.
\bibitem{carlson} R. F. Carlson, Atomic Data and
Nuclear Data tables {\bf 63} (1996) 93.
\bibitem{czyz} W. Czyz and L. C. Maximon, Ann. Phys. (N.Y.) {\bf 52} (1969) 59.
\bibitem{c.m.correl} V. Franco and G. K. Varma, Phys. Rev. C {\bf 18} 
(1978) 349.  
\bibitem{skoyb} Y. Suzuki, T. Kido, Y. Ogawa, K. Yabana and D. Baye, 
Nucl. Phys. A {\bf 567} (1994) 957.
\bibitem{cooper} E. D. Cooper, S. Hama, B. C. Clark and 
R. L. Mercer, Phys. Rev. C {\bf 47} (1993) 297.
\end{thebibliography}
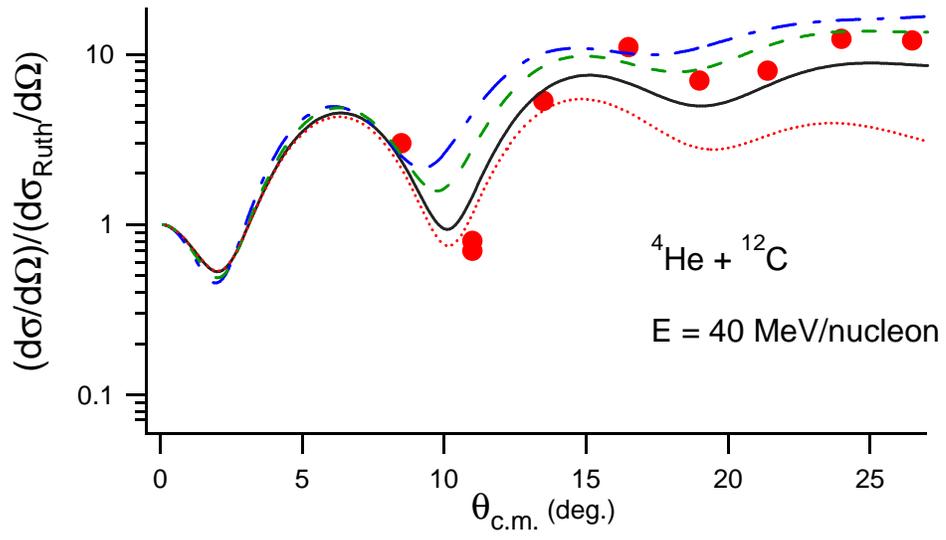
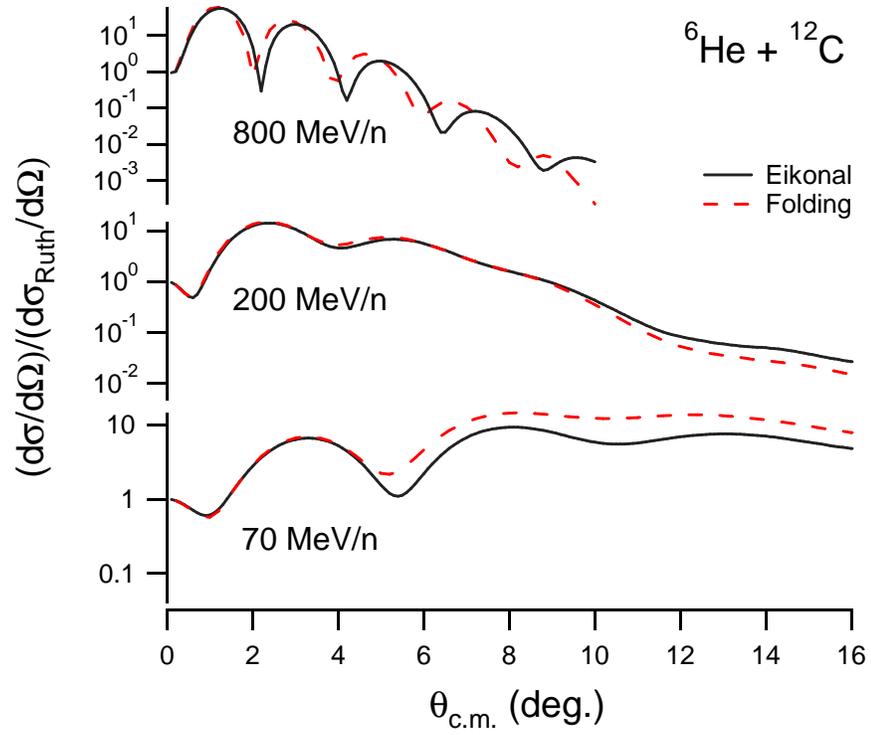
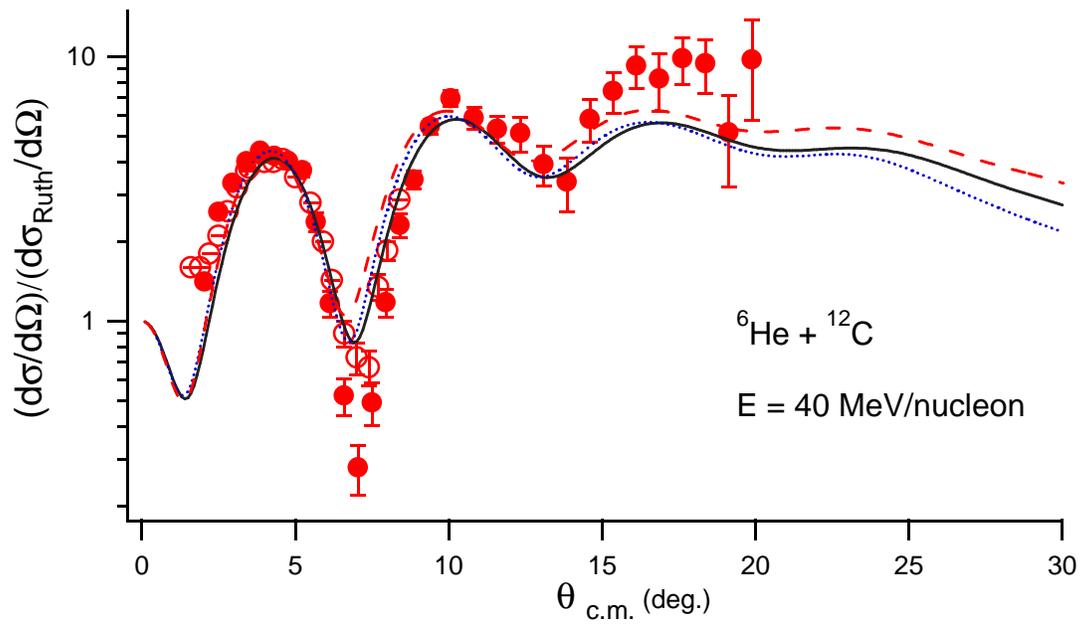
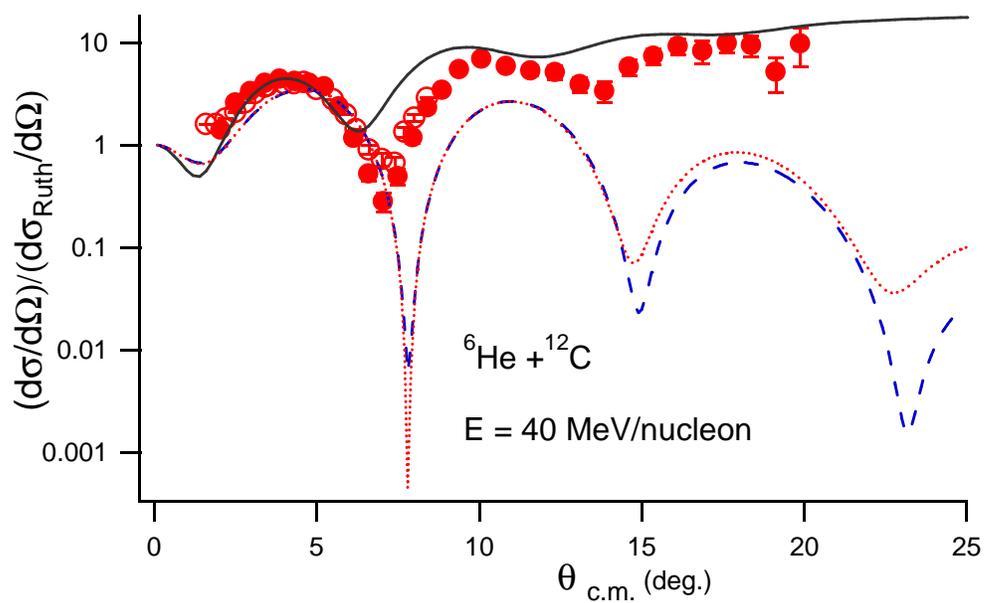
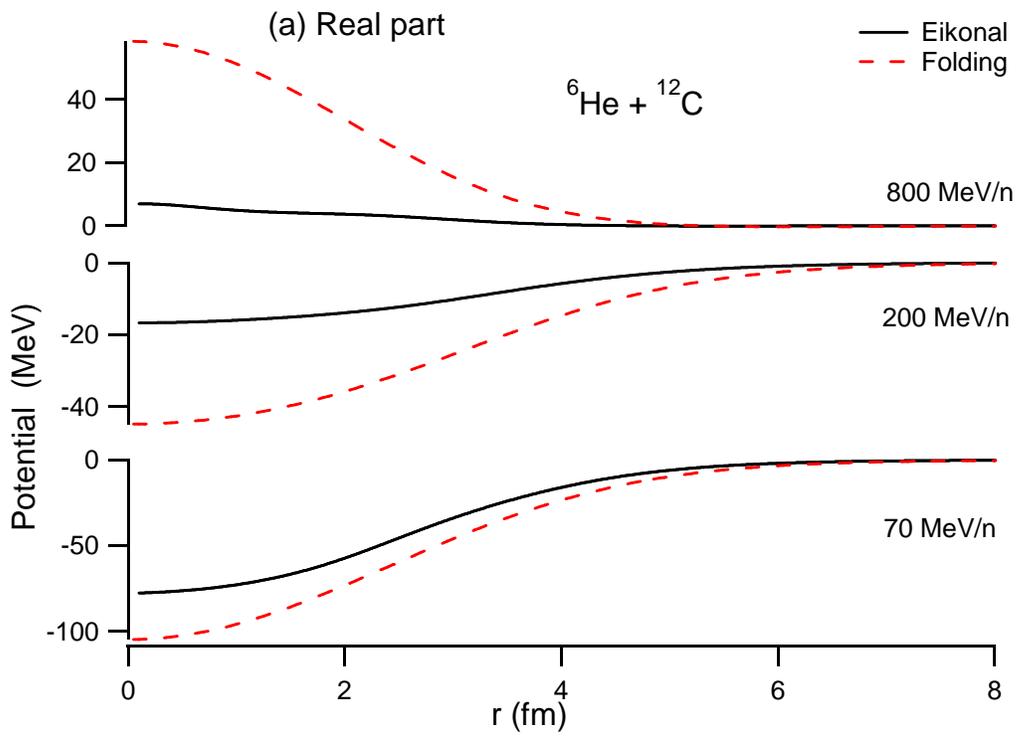
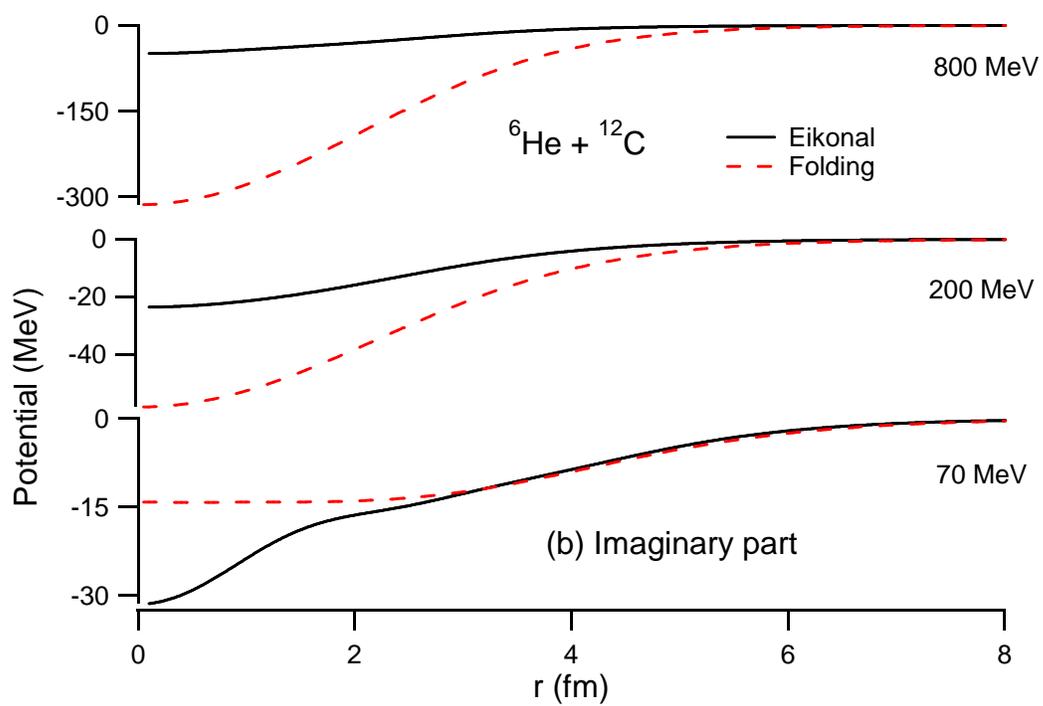
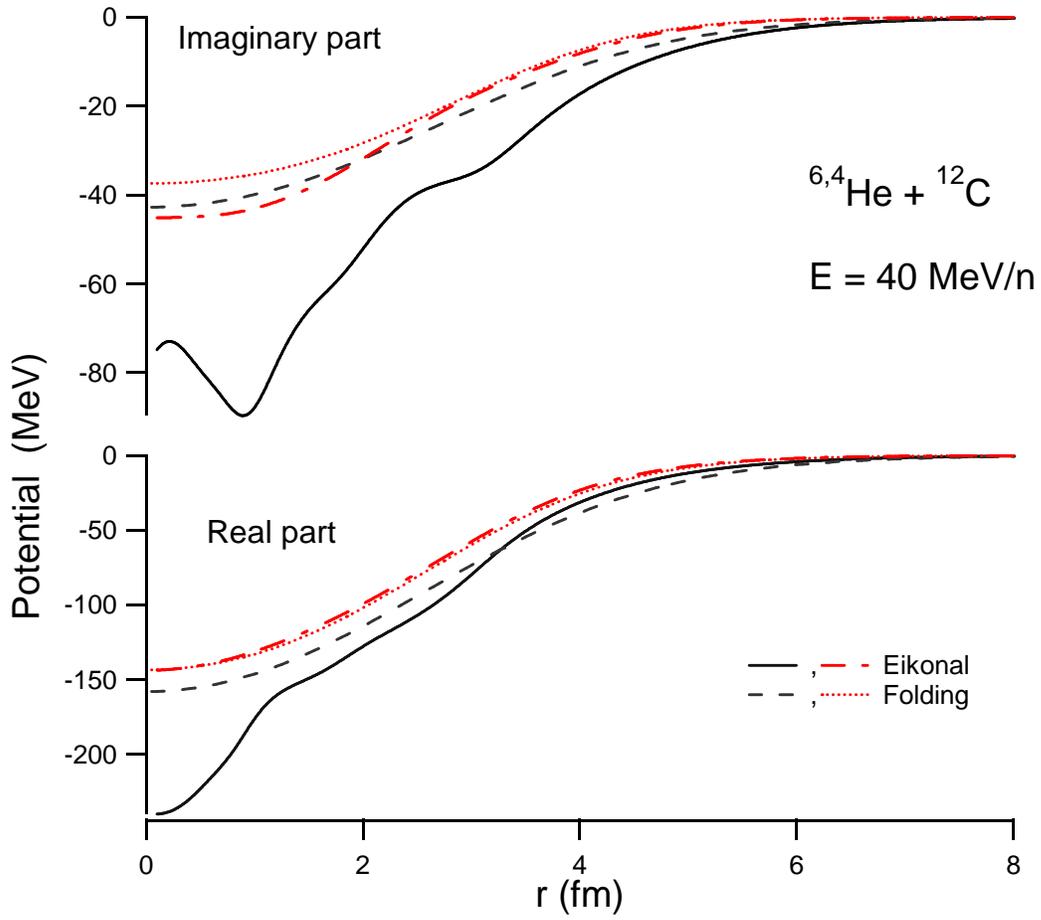
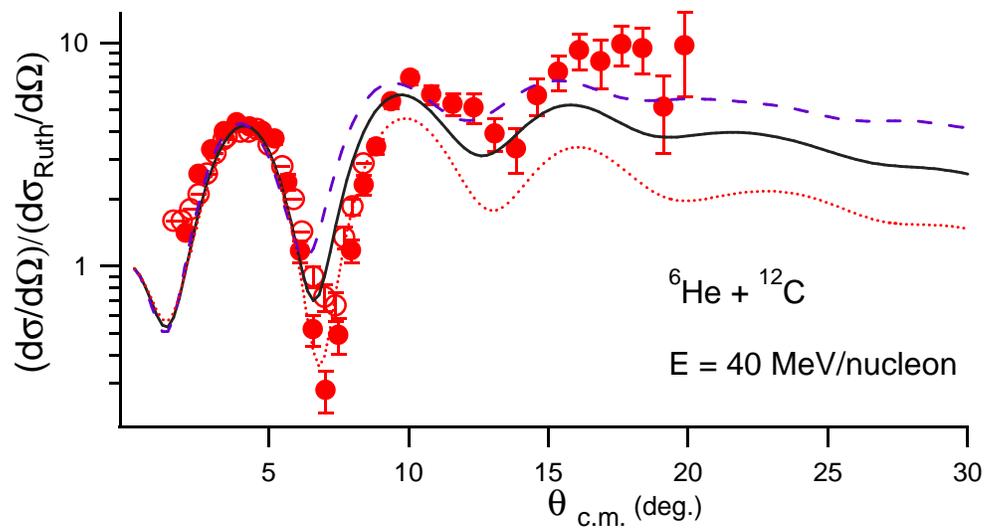
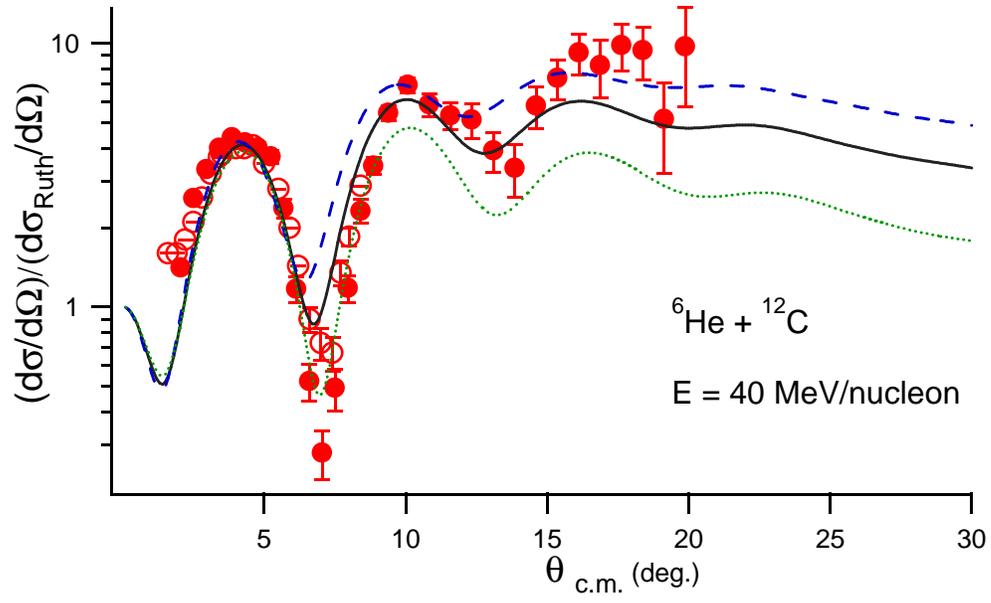
\end{document}